\documentclass[journal]{IEEEtran}

\usepackage[noadjust]{cite}
\usepackage{graphicx,epstopdf}
\usepackage[table]{xcolor}
\usepackage{amsmath,amsfonts,amssymb,amsthm,mathtools}
\usepackage{tikz,tikz-3dplot,tikzinclude,url,array}

\usetikzlibrary{intersections,shapes,arrows,positioning}
\usetikzlibrary{shapes.geometric,calc,patterns}

\usepackage{multirow}
\usepackage{booktabs}

\usepackage{hyperref}
\hypersetup{
	colorlinks,
	linkcolor={violet!50!black},
	citecolor={purple!50!black},
	urlcolor={blue!50!black}
}

\newcommand{\mysty}[1]{\textcolor{green!50!black}{\textbf{#1}}}
\begin{document}

\title{Airfoil's Aerodynamic Coefficients Prediction using Artificial Neural Network}


\author{\IEEEauthorblockN{Hassan Moin, Hafiz Zeeshan Iqbal Khan, Surrayya Mobeen and Jamshed Riaz}\\
\IEEEauthorblockA{Department of Aeronautics \& Astronautics,\\ Institute of Space Technology, Islamabad, 44000, Pakistan.}\vspace{-0.75cm}}


\maketitle

\begin{abstract}
Figuring out the right airfoil is a crucial step in the preliminary stage of any aerial vehicle design, as its shape directly affects the overall aerodynamic characteristics of the aircraft or rotorcraft. Besides being a measure of performance, the aerodynamic coefficients are used to design additional subsystems such as a flight control system, or predict complex dynamic phenomena such as aeroelastic instability. The coefficients in question can either be obtained experimentally through wind tunnel testing or, depending upon the accuracy requirements, by numerically simulating the underlying fundamental equations of fluid dynamics. In this paper, the feasibility of applying Artificial Neural Networks (ANNs) to estimate the aerodynamic coefficients of differing airfoil geometries at varying Angle of Attack, Mach and Reynolds number is investigated. The ANNs are computational entities that have the ability to learn highly nonlinear spatial and temporal patterns. Therefore, they are increasingly being used to approximate complex real-world phenomenon. However, despite their significant breakthrough in the past few years, ANNs’ spreading in the field of Computational Fluid Dynamics (CFD) is fairly recent, and many applications within this field remain unexplored. This study thus compares different network architectures and training datasets in an attempt to gain insight as to how the network perceives the given airfoil geometries, while producing an acceptable neuronal model for faster and easier prediction of lift, drag and moment coefficients in steady state, incompressible flow regimes. This data-driven method produces sufficiently accurate results, with the added benefit of saving high computational and experimental costs.
\end{abstract}


\section{Introduction}
Airfoil design is a major facet of aircraft aerodynamics. An airfoil-shaped body moving through air produces an aerodynamic force and moment. The component of this force perpendicular to the direction of motion is called lift, while the component parallel to the direction of motion is called drag. During any aero-vehicle design process, an aerospace engineer will use the airfoil geometry to calculate the preliminary lift, drag and moment coefficients, center of pressure, aerodynamic center and other pertinent parameters. These calculations are then extrapolated to derive the aerodynamic characteristics of a finite wing in order to assess its performance. Hence, numerical search for the optimum shape of airfoil/wing geometry is of immense significance to aircraft engineers.

Due to recent advancements in computational engineering, various methods are being employed to compute aerodynamic coefficients and optimum airfoil geometry in a more efficient manner. Moreover, rise in the field of computer science and Artificial Intelligence (AI) based techniques have led to development of highly reliable data-driven models of physical systems including, but not limited to, aerodynamic applications. These surrogate models have an added advantage of predicting solutions to nonlinear problems efficiently and accurately, which is why there has been a recent surge in interest amongst the scientific community for developing such formidable models to tackle real-world problems.

CFD methods, which are computationally intensive, rely profoundly on physical laws. However, their implementation is impractical in certain situations (e.g. real-time control purposes) since they require time consuming simulations. In contrast, models that rely heavily on data, such as ANNs, can do pure input-output mapping rapidly, without a priori knowledge of the physical system. ANNs can therefore be ideal for flight control applications, and can be used effectively for identification and control of dynamical systems \cite{Narendra1990,Hunt1992,Calise1998}.

By now, numerous applications of AI based techniques exist in the field of fluid mechanics. This includes steady-state, incompressible flows. Multiple studies have been carried out to predict flow fields and aerodynamic force coefficients of airfoils using a specialized class of neural networks that are most commonly applied to the analysis of visual imagery. These methods are known as deep learning techniques and utilize convolution layers to extract unique features from images and use them for training. Zhang et al. \cite{Zhang2018} trained multiple of these Convolutional Neural Networks (CNNs) to learn the lift coefficients of different airfoil shapes at varying Angles of Attack (AoA), Mach and Reynolds numbers; while Sekar et al. \cite{Sekar2019} used CNNs to approximate the flow field over an airfoil as a function of airfoil geometry, Reynolds number, and AoA without directly solving the Navier Stokes equations. Bhatnagar et al. \cite{Bhatnagar2019} predicted the velocity and pressure fields using CNNs and Chen et al. \cite{Chen2020} generated a dataset of composite airfoil images using flow-condition convolution in order to predict aerodynamic coefficients. Hui et al. \cite{Hui2020} used CNNs to predict the pressure distribution around airfoils, while Guo et al. \cite{Guo2016} used them to predict non-uniform steady laminar flow in a 2D or 3D domain.

On the other hand, simple ANN architectures have commonly been used for the inverse airfoil design problem. Rai and Madavan used pressure distribution \cite{Rai2001} along with 15 other design variables \cite{Rai2000} at the input of ANNs to design turbomachinery airfoils. Huang et al. \cite{Huang1994} designed and evaluated the Eppler method using ANNs. The Eppler method is a multipoint inverse airfoil design method that anticipates the airfoil performance at different conditions by specifying information on the airfoil's velocity distribution. In their study, the authors predicted lift coefficients with notable accuracy, however, high error was observed between the predicted and actual drag coefficients. Khurana et al. \cite{Khurana2008} used a swarm-based approach as a search agent with an ANN on PARSEC (a common method for airfoil parameterization) airfoils in order to optimize their shapes. Greenman and Roth \cite{Greenman1999} optimized the high lift performance of a multi-element airfoil using neural networks. Hacioglu \cite{Hacioglu2007} used genetic algorithm in conjunction with a trained ANN to search the airfoil design space with improved algorithmic speed. Sekar et al. \cite{Sekar2019a} used a CNN for the inverse airfoil design problem, while Sun et al. \cite{Sun2015} used ANNs for inverse design of both airfoils and wings.

Moreover, Li et al. \cite{Li2020} proposed a new sampling method for airfoils and wings, based upon a Deep Convolutional-Generative Adversarial Network (DC-GAN). Xu et al. \cite{Xu2019} used neural networks to optimize the design of airfoil in presence of buffeting phenomenon. Dupuis et al. \cite{Dupuis2018} presented an innovative approach for the prediction of steady turbulent aerodynamic fields using neural networks, whereas Singh et al. \cite{Singh2017} reconstructed models through neural networks to better predict coefficient of lift and flow separation over airfoils.

In this study, we discuss the implementation of a simple yet novel idea, whereby we estimate aerodynamic coefficients of different airfoils by ensuring that the ANN learns the airfoil geometric design space using normalized 2D coordinates instead of airfoil design parameters.

\section{Preliminaries}
\subsection{Airfoil \& its Aerodynamics}
\begin{figure}
  \centering
  \includegraphics[width=\linewidth]{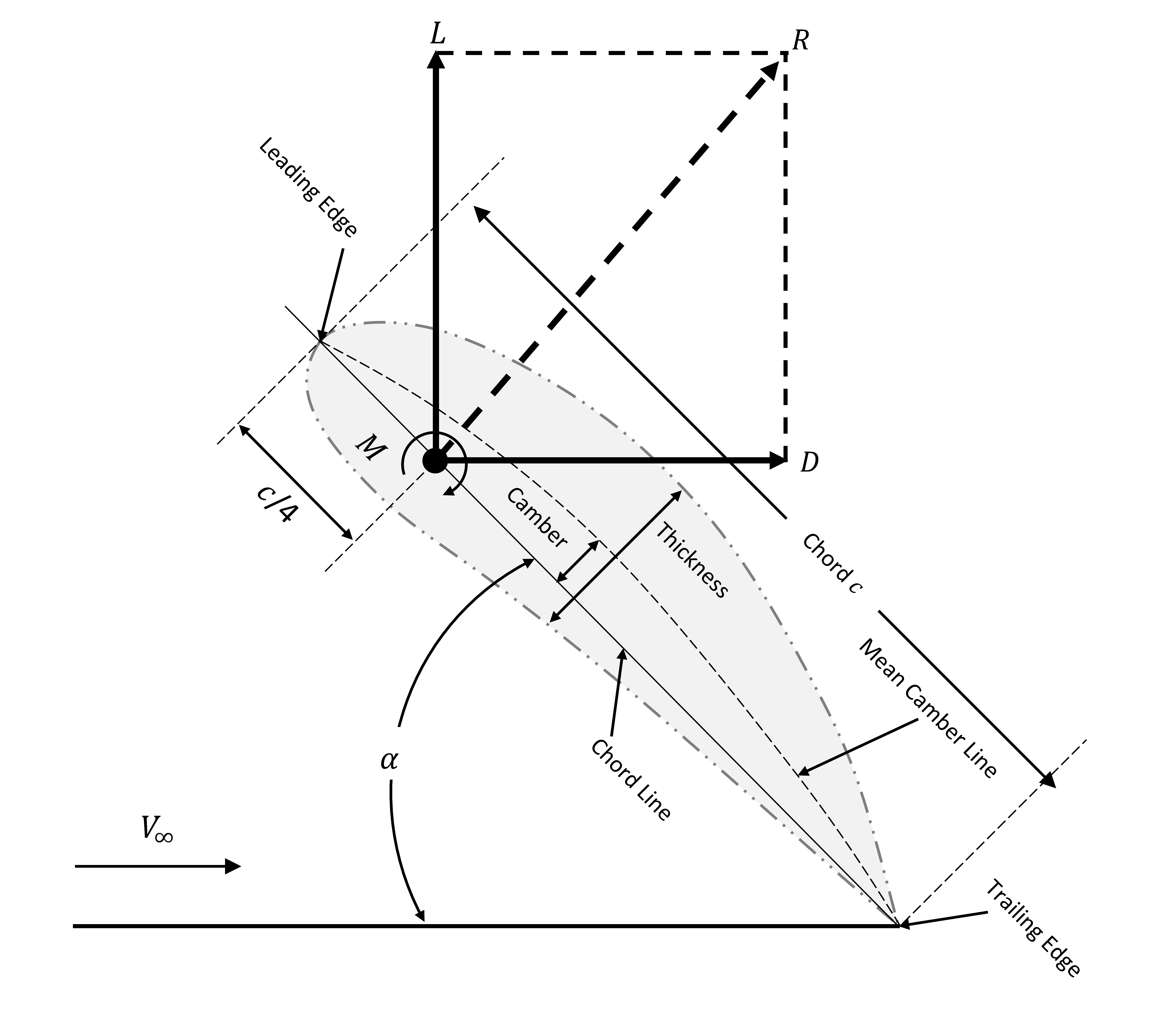}
  \caption{Airfoil Nomenclature}\label{fig:airfoil}
\end{figure}

An airfoil is described as the cross-section of a wing. As mentioned previously, the geometric shape of the airfoil determines some very important aerodynamic coefficients that in turn determine the design and performance of the entire aircraft. Figure \ref{fig:airfoil} illustrates the common nomenclature related to airfoils. These terminologies are described as follows:

\begin{itemize}
  \item \emph{Mean Camber Line}: The line connecting mean points halfway between the upper and lower surfaces of an airfoil.
  \item \emph{Leading Edge Radius}: The radius of the circle centered at a point on the mean camber line, and tangent to the upper and lower surfaces at the forward-most edge of an airfoil.
  \item \emph{Leading \& Trailing Edge}: The forward-most and the rear-most edge of an airfoil, respectively.
  \item \emph{Chord Line}: A straight line having length c, that connects the forward-most and rear-most points of an airfoil.
  \item \emph{Camber}: The maximum distance between mean camber line and chord line measured perpendicularly.
  \item \emph{Thickness}: Maximum distance between the upper and lower surfaces of an airfoil.
  \item \emph{Angle of Attack ($\alpha$)}: The angle that the relative wind ($V_\infty$) makes with the chord line.
  \item \emph{Lift ($L$)}: The component of aerodynamic force ($R$) perpendicular to the relative wind ($V_\infty$), that enables the aircraft to fly.
  \item \emph{Drag ($D$)}: The component of aerodynamic force ($R$) parallel to the relative wind ($V_\infty$), that opposes the forward motion.
  \item \emph{Moment ($M$)}: Tendency of the airfoil to rotate about a fixed point due to pressure and shear stress distributions.
\end{itemize}

An airfoil’s mean camber line, location of maximum camber, and to a lesser extent, thickness distribution and the location of maximum thickness essentially controls its lift, drag and moment characteristics.

The two forces and one moment are normally converted into dimensionless coefficients for ease:

\[
C_L = \frac{L}{qc}, \qquad C_D = \frac{D}{qc}, \qquad C_m = \frac{M}{qc^2}
\]

where $q=\frac{1}{2}\rho V_\infty^2$ is the dynamic pressure.

A large volume of experimental airfoil data was recorded over the years by the National Advisory Committee for Aeronautics (NACA), which later on became a part of National Aeronautics and Space Administration (NASA) in 1958. The airfoils were characterized into multiple series, for example: NACA 4- and 5-digit series. These two series are considered in this work. NACA 4-digit series is defined by the following profile \cite{Eastman1933,Abbott1959}:

\begin{enumerate}
  \item First digit describes the maximum camber as percentage of the chord.
  \item Second digit describes the distance of maximum camber from the airfoil leading edge in tenths of the chord.
  \item Last two digits describe maximum thickness of the airfoil in form of percentage of the chord.
\end{enumerate}

The NACA five-digit series describes more complex airfoil shapes. Its format is LPSTT \cite{Abbott1959,Jacobs1936}, where:

\begin{enumerate}
  \item L: a single digit representing the theoretical optimal lift coefficient at ideal angle of attack
  \item P: a single digit for the $x$-coordinate of the point of maximum camber
  \item S: a single digit indicating whether the camber is simple (S = 0) or reflex (S = 1)
  \item TT: the maximum thickness in percent of chord, as in a four-digit NACA airfoil code.
\end{enumerate}

\subsection{Artificial Neural Networks (ANNs)}
Artificial Neural Networks are self-learning entities that are designed to simulate the way a human brain analyzes and processes information via a web of interconnected neurons. They are massive parallel computation paradigms consisting of simple processing elements that can solve problems which are deemed complex in nature by human or statistical standards, given enough learning examples are available. Since the past few years, ANNs are being increasingly used in a variety of applications due to their adaptive nature and excellent function approximation capability.

The ANN architecture consists of multiple layers of artificial neurons. Each neuron computes a static nonlinear activation function once it receives a set of numerical information from the previous layer at its input. Hence, the output of layer $L$ is defined as the activation $a^L$, which is a nonlinear transformation of the weighted sum of outputs $a^{L-1}$ of the $(L-1)$th layer, including a bias term $b^{L-1}$:
\begin{equation}
\begin{split}
   z^L &= \Theta^L a^{L-1} + b^{L-1} \\
   a^L &= g(z^L)
\end{split}, \qquad \forall\,L \in [1,N_L]
\end{equation}
If the network has $s_{L-1}$ neurons in layer $L-1$, and $s_L$ neurons in layer $L$, $\Theta^{L-1}$ is defined as the weight matrix having $s_L\times(s_{L-1}+1)$ dimensions. This weight matrix can be intuitively thought of as a structure that maps the output values of layer $L-1$ to layer $L$. $g(z^L)$ is defined as a vector of outputs of individual neurons based on a nonlinear activation functions. The final output of the nested functions showcasing the successive connections of an $N_L$-layered ANN is $a^{N_L}$, where $a^0$ is the input vector.

The \emph{learning} part of any ANN is done through its backpropagation algorithm. Once the forward pass computes values from inputs to outputs, a loss function is used to calculate the error between the actual and the predicted value. The sensitivity of the cost with respect to each weight is then computed using the backward pass, which is considered as a recursive application of the chain rule along a computational graph. The backpropagation algorithm minimizes the loss function using common optimization algorithms such as stochastic gradient descent or the ADAM optimizer.

\section{Methodology}
This section outlines the process of obtaining, cleaning and transforming raw data prior to feeding it to the learning algorithm. It is an essential step that often involves making corrections to data, reformatting data, and combining datasets to enrich the overall training data.

\subsection{Data Preparation}
\begin{figure*}
  \centering
  \includegraphics[width=\linewidth]{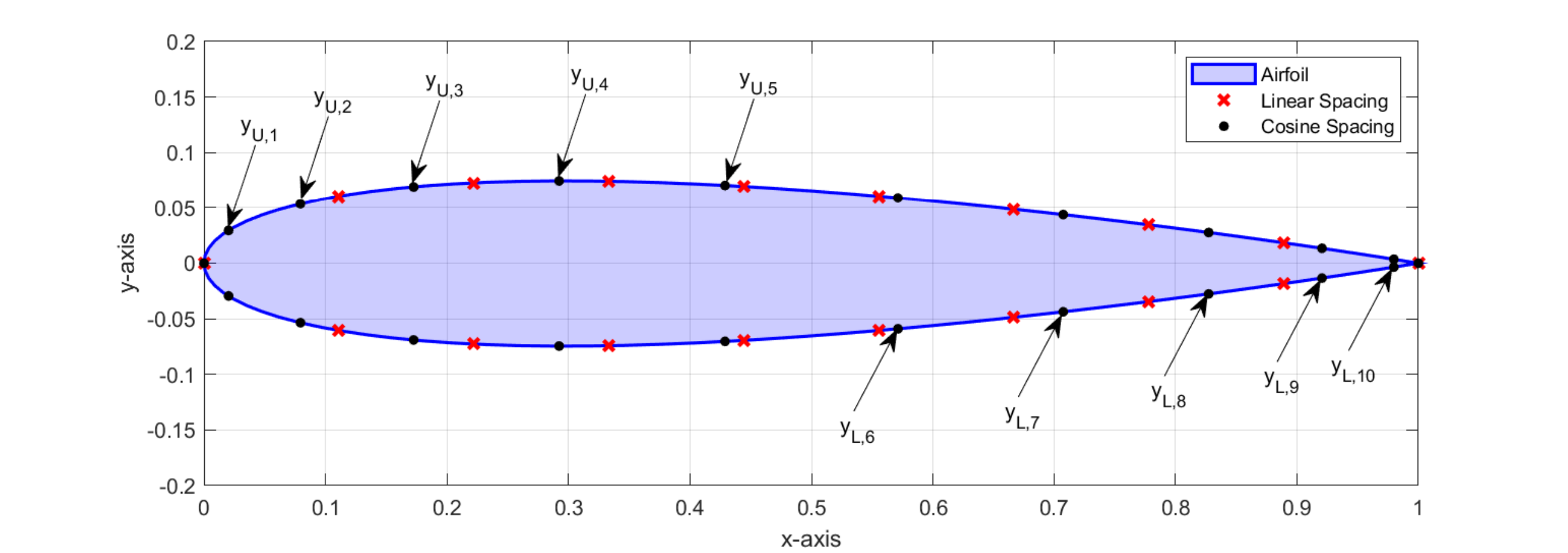}
  \caption{Spacing at $\mathcal{N}=10$}\label{fig:CosineSpacing}
\end{figure*}

Due to their wide range of applications, a dataset of NACA 4- and 5-digits airfoils is generated in \emph{Javafoil} using macros. Depending on the shape, each airfoil is discretized at 101 \emph{cosine-spaced} points (normalized to unit chord length), in order to generate smooth upper and lower surfaces. For each airfoil, the corresponding lift ($C_L$), drag ($C_D$) and moment ($C_m$) coefficients are also obtained at different AoAs, Reynolds and Mach numbers using the same software. Next, these coordinate points are interpolated at fixed $\mathcal{N}$ locations on the $x$-axis, which are distributed via cosine spacing (see Figure \ref{fig:CosineSpacing}), where $y_{U,k}$ and $y_{L,k}$, for all $k\in[1,\mathcal{N}]$, are upper and lower surface points, respectively. This method of spacing is normally used to capture the leading and trailing edge shapes by having denser points around these areas as compared to center. Moreover, it must be noted that the leading and trailing edges are kept fixed at $(0,0)$ and $(1,0)$ respectively. However, these two points are ignored in our dataset as they are constant in every training example.

Based on these design specifications, six different datasets were created (three for each series). Their subsequent names, along with the airfoil parameters and flight conditions are shown in Table \ref{tab:DataRanges}.

\renewcommand\arraystretch{1.3}
\begin{table*}
\centering
\caption{Ranges for Airfoil Parameters \& Flight Conditions}
\label{tab:DataRanges}
\arrayrulecolor{black}
\begin{tabular}{clcccccc}
\toprule
\multicolumn{2}{c}{No. of Coordinate Points ($\mathcal{N}$)}  & \multicolumn{3}{c}{NACA 4-Digit Series} & \multicolumn{3}{c}{NACA 5-Digit Series}                       \\
\cmidrule(rl){1-2}\cmidrule(rl){3-5}\cmidrule(rl){6-8}
\multicolumn{2}{c}{5}                                                                                                                                         & \multicolumn{3}{c}{$\mathcal{D}^4_{05}$} & \multicolumn{3}{c}{$\mathcal{D}^5_{05}$}  \\
\multicolumn{2}{c}{10}                                                                                                                                        & \multicolumn{3}{c}{$\mathcal{D}^4_{10}$} & \multicolumn{3}{c}{$\mathcal{D}^5_{10}$}  \\
\multicolumn{2}{c}{15}                                                                                                                                        & \multicolumn{3}{c}{$\mathcal{D}^4_{15}$} & \multicolumn{3}{c}{$\mathcal{D}^5_{15}$}  \\
\midrule
\multicolumn{2}{l}{~}                                                                                                       & Min    & Max    & Step   & Min    & Max    & Step                              \\
\cmidrule(rl){1-2}\cmidrule(rl){3-5}\cmidrule(rl){6-8}
& Thickness (\%)                     & 5      & 35     & 5                                         & 5                                  & 35     & 5          \\
& Location of Max. Thickness (\%) & \multicolumn{3}{c}{30 (fixed)}                               & \multicolumn{3}{c}{30 (fixed)}                                \\
& Max. Camber (\%)                & 0      & 9      & 1                                          & \multicolumn{3}{c}{Not Applicable}                                        \\
& Design Lift Coefficient            & \multicolumn{3}{c}{Not Applicable}                                       & 0      & 1.8    & 0.2                                         \\
& Location of Max. Camber (\%)    & 5      & 75     & 10                                         & 5      & 75     & 10  \\
\multirow{-6}{2cm}{Airfoil Parameters} & Reflexed Trailing Edge & \multicolumn{3}{c}{Not Applicable}  & \multicolumn{3}{c}{Both Standard \& Reflexed}          \\
\cmidrule(rl){1-2}
& Angle of Attack (deg)              & -10    & 10     & 1                                          & -10    & 10     & 1                                           \\
& Reynolds No.                       & $1\times10^5$ & $5\times10^5$ & $1\times10^5$                                     & $1\times10^5$ & $5\times10^5$ & $1\times10^5$                                      \\
\multirow{-3}{2cm}{Flight Conditions} & Mach No.  & 0.1    & 0.3    & 0.1                                        & 0.1    & 0.3    & 0.1                                         \\
\bottomrule
\end{tabular}
\arrayrulecolor{black}
\end{table*}
\renewcommand\arraystretch{1.0}

Based on the chosen parameter space, we obtain a set of 560 airfoils for NACA 4-digit series and 1120 airfoils for NACA 5-digit series, each having their aerodynamic characteristics solved at a combination of 315 different flight conditions. That makes a total of 176400 and 352800 samples in each 4- and 5-digit dataset, respectively.

\subsection{Data Preprocessing}
Reformatting, cleaning and data processing are the next three major steps undertaken before feeding the samples into an ANN. A script was written in Python which imports all files, stacking them together in the required format, while removing unnecessary data. The final array obtained had $N_s\times(2\mathcal{N}+6)$ dimensions, where $N_s$ is the number of row-wise samples. The first $\mathcal{N}$ columns are comprised of $y$-coordinates of the upper surface at fixed $x$ locations and the next $\mathcal{N}$ columns consist of $y$-coordinates of the lower surface at the same $x$ locations. The last six columns consist of AoA, Reynold and Mach number, $C_L$, $C_D$ and $C_m$ values, respectively. Moreover, few airfoils of NACA 5-digit series were found out not to have their trailing edges on x-axis. Upon further inspection, few irregularities were observed in certain samples due to limitations of \emph{Javafoil}. The software was unable to compute one or more coefficients of airfoils that had a very high camber, especially if it existed near one of the edges. These erroneous airfoils and data samples were removed entirely from each dataset. Hence, we were left with 171431 and 283244 data samples in 4- and 5-digit datasets, respectively.

\begin{figure}
  \centering
  \includegraphics[width=\linewidth]{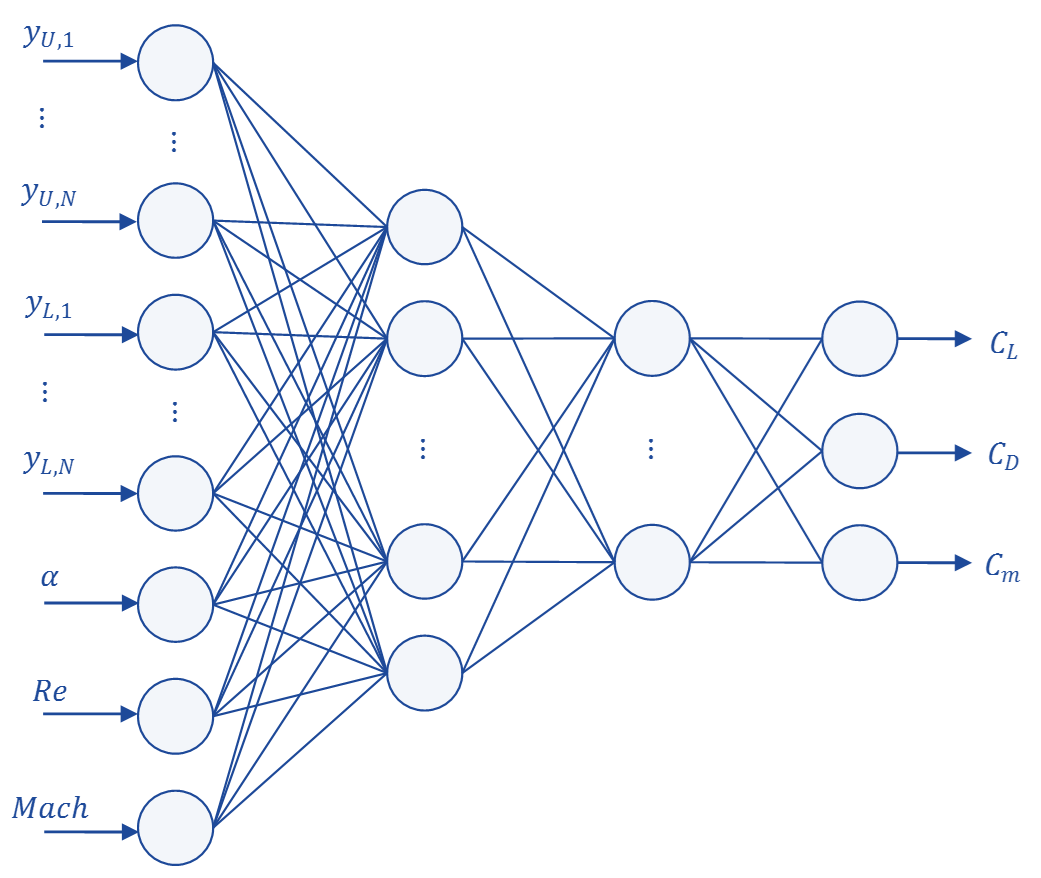}
  \caption{Artificial Neural Network}\label{fig:ANN}
\end{figure}

Next, the samples in each dataset were randomly shuffled in order to improve the overall model quality, and enhance its predictive performance. Once shuffled, they were then split into training, validation and test sets with a ratio of 70:15:15. The input columns (features) were separated in each set from the output columns (labels). Finally, all ($2\mathcal{N}+3$) features in the training set were standardized by removing their means and scaling to unit variance. Likewise, the features for validation and test sets were also normalized by the same parameters. A model showcasing the input and outputs of the neural network is illustrated in Figure \ref{fig:ANN}.

\subsection{Network Architecture}
All neural networks were designed using Keras API in Python. The hardware that was used to train the network had the following specifications: HP Probook 450 G4, Microsoft Windows 10 Pro 64-bit (OS) 10.0.18363 (Build), Intel Ci7-7500U (Processor) @ 2.70 GHz, 2904 MHz, 2 Cores, 16 GB (RAM), 256 SSD + 1 TB HDD (Hard Drives) and 2 GB NVIDIA GeForce 930MX (Graphics Card). Adam optimizer was utilized as the optimization scheme, where the learning rate was initially kept as 0.0005. The first and second moments were maintained at 0.9 and 0.999 respectively. Mean Squared Error (MSE) was kept as the loss function, while Root Mean Squared Error (RMSE) and R-squared values were kept as metrics to measure the performance of the networks. Rectified Linear Unit (ReLU) was used as the activation function in all layers due to its computational simplicity and effectiveness in handling of vanishing gradients. Every network was trained for 50 epochs with the data being fed to the network in multiple batches of 128 samples. Validation losses were also monitored simultaneously, and the learning rate was automatically reduced by a factor of 10\% if the validation loss did not improve after a ‘patience’ period of 5 epochs.

\section{Results \& Discussion}
A total of 20 runs were conducted for each case outlined in Tables \ref{tab:Layers}, \ref{tab:Neurons}, \ref{tab:Points}, and \ref{tab:Dataset}. The average losses across the train, validation and test datasets were similar in each run, indicating that the models had generalized well within the first 10 epochs. Therefore, RMSE and $R^2$ values of models with the best average loss on test sets, across multiple runs, were recorded in the aforementioned tables.

The goal for conducting these iterative studies was to select a computationally inexpensive architecture exhibiting the greatest acceptable performance over the three parameters of interest (i.e. $C_L$, $C_D$, $C_m$). Normally, extremely wide, shallow networks are very good at memorization, but they are not so good at generalization. In contrast, deep networks are much better at generalizing, as they can capture nonlinearities and can learn features at various levels of abstraction. Thus, a tradeoff between the number of layers and neurons was required.

\renewcommand\arraystretch{1.3}
\begin{table*}
\centering
\caption{Network Performance due to Multiple Hidden Layers (Dataset: $\mathcal{D}^4_{10}$)}\label{tab:Layers}
\arrayrulecolor{black}
\begin{tabular}{cccccccc}
\toprule
\multirow{2}{*}{Case No.}  & \multirow{2}{*}{Network Architecture} & \multicolumn{3}{c}{RMSE} & \multicolumn{3}{c}{$R^2$}   \\
& & $C_L$ & $C_D$ & $C_m$ & $C_L$ & $C_D$  & $C_m$ \\
\cmidrule(rl){1-1}\cmidrule(rl){2-2}\cmidrule(rl){3-5}\cmidrule(rl){6-8}
1 & 64, 3        & 0.038998 & 0.009348 & 0.016889 & 0.997735 & 0.835575 & 0.969677 \\
2 & 64, 32, 3    & 0.032737 & 0.009866 & 0.014702 & 0.998404 & 0.816859 & 0.977019 \\
\mysty{3} & \mysty{64, 32, 16, 3} & \mysty{0.027694} & \mysty{0.009906} & \mysty{0.013390} & \mysty{0.998858} & \mysty{0.815375} & \mysty{0.980938}  \\
4 & 64, 32, 16, 8, 3 & 0.029456 & 0.009853 & 0.013259 & 0.998708 & 0.817350 & 0.981309 \\
\bottomrule
\end{tabular}
\arrayrulecolor{black}
\end{table*}
\renewcommand\arraystretch{1.0}

\renewcommand\arraystretch{1.3}
\begin{table*}
\centering
\caption{Network Performance due to Increasing Neurons (Dataset: $\mathcal{D}^4_{10}$)}\label{tab:Neurons}
\arrayrulecolor{black}
\begin{tabular}{cccccccc}
\toprule
\multirow{2}{*}{Case No.} & \multirow{2}{*}{Network Architecture} & \multicolumn{3}{c}{RMSE} & \multicolumn{3}{c}{$R^2$}   \\
& & $C_L$ & $C_D$ & $C_m$ & $C_L$ & $C_D$  & $C_m$ \\
\cmidrule(rl){1-1}\cmidrule(rl){2-2}\cmidrule(rl){3-5}\cmidrule(rl){6-8}
1 & 64, 32, 16, 3 & 0.027694 & 0.009906 & 0.013390 & 0.998858 & 0.815375 & 0.980938 \\
2 & 128, 64, 32, 3 & 0.025191 & 0.008612 & 0.011476 & 0.999055 & 0.860445 & 0.985998 \\
3 & 256, 128, 64, 3 & 0.020300 & 0.007233 & 0.008163 & 0.999386 & 0.901568 & 0.992916 \\
\mysty{4} & \mysty{512, 256, 128, 3} & \mysty{0.019227} & \mysty{0.006776} & \mysty{0.007532} & \mysty{0.999449} & \mysty{0.913612} & \mysty{0.993970} \\
5 & 1024, 512, 256, 3 & 0.016462 & 0.006037 & 0.005809 & 0.999596 & 0.931430 & 0.996412 \\ \bottomrule
\end{tabular}
\arrayrulecolor{black}
\end{table*}
\renewcommand\arraystretch{1.0}

\renewcommand\arraystretch{1.3}
\begin{table*}
\centering
\caption{Network Performance due to Varying Geometric Information (Architecture: [512, 256, 128, 3])} \label{tab:Points}
\arrayrulecolor{black}
\begin{tabular}{cccccccc}
\toprule
\multirow{2}{*}{Case No.}  & \multirow{2}{*}{Dataset} & \multicolumn{3}{c}{RMSE} & \multicolumn{3}{c}{$R^2$}   \\
& & $C_L$ & $C_D$ & $C_m$ & $C_L$ & $C_D$  & $C_m$ \\
\cmidrule(rl){1-1}\cmidrule(rl){2-2}\cmidrule(rl){3-5}\cmidrule(rl){6-8}
1 & $\mathcal{D}^4_{05}$ & 0.022022 & 0.006902 & 0.007382 & 0.999278 & 0.910371 & 0.994206 \\
\mysty{2} & \mysty{$\mathcal{D}^4_{10}$} & \mysty{0.017082} & \mysty{0.006344} & \mysty{0.006378} & \mysty{0.999566} & \mysty{0.924274} & \mysty{0.995674}  \\
3 & $\mathcal{D}^4_{15}$ & 0.018210 & 0.006484 & 0.006862 & 0.999056 & 0.920897 & 0.994994 \\
\bottomrule
\end{tabular}
\arrayrulecolor{black}
\end{table*}
\renewcommand\arraystretch{1.0}

\renewcommand\arraystretch{1.3}
\begin{table*}
\centering
\caption{Network Performance on Different Datasets (Architecture: [512, 256, 128, 3])}
\label{tab:Dataset}
\arrayrulecolor{black}
\begin{tabular}{ccccccccc}
\toprule
\multirow{2}{*}{Case No.}  & \multicolumn{2}{c}{Dataset} & \multicolumn{3}{c}{RMSE} & \multicolumn{3}{c}{$R^2$}   \\
& Train & Test & $C_L$ & $C_D$ & $C_m$ & $C_L$ & $C_D$  & $C_m$ \\
\cmidrule(rl){1-1}\cmidrule(rl){2-3}\cmidrule(rl){4-6}\cmidrule(rl){7-9}
1 & $\mathcal{D}^4_{10}$ & $\mathcal{D}^5_{10}$ & 0.131897 & 0.015882 & 0.021380 & 0.968168 & 0.678719 & 0.617477 \\
2 & $\mathcal{D}^5_{10}$ & $\mathcal{D}^4_{10}$ & 0.094408 & 0.012840 & 0.032633 & 0.986710 & 0.706630 & 0.885493 \\
\mysty{3} & \mysty{$\mathcal{D}^4_{10} \cup \mathcal{D}^5_{10}$} & \mysty{$\mathcal{D}^4_{10} \cup \mathcal{D}^5_{10}$} & \mysty{0.014061} & \mysty{0.007877} & \mysty{0.004215} & \mysty{0.999681} & \mysty{0.909185} & \mysty{0.996986} \\
\bottomrule
\end{tabular}
\arrayrulecolor{black}
\end{table*}
\renewcommand\arraystretch{1.0}

The first set of cases was implemented to establish the depth of our proposed network architecture. $\mathcal{D}^4_{10}$ dataset was arbitrarily chosen, and the architecture was kept simple, with the number of neurons halving in every subsequent hidden layer. The last layer, commonly known as the output layer, consisted of three neurons encapsulating the estimated value of each predicted variable. As seen from Table \ref{tab:Layers}, the overall performance improved as we increased the architecture from a two-layered network to a four-layered network. However, the average performance slightly deteriorated on the five-layered network.

The next set of cases was implemented to determine the overall width of our ANN architecture. We selected a four-layered network based on the results of Table \ref{tab:Layers}, and doubled the neurons in each hidden layer as indicated in Table \ref{tab:Neurons}. It was observed that as the number of neurons increased; the average performance of the networks increased as well, but at the expense of high computational requirements. The difference in performance of case 4 and 5 in Table \ref{tab:Neurons} was not significant enough to justify the increase in computational complexity. Therefore, for the scope of this study, the results of case 4 were deemed as sufficient.

Moreover, the network performance was studied by varying the \emph{`quality'} of geometric information (provided as input data). Unlike in CFD analysis, where geometric profiles and mesh sizes play a huge role in identifying the aerodynamic characteristics, it was also explored if the network would be able to identify these characteristics or not, given a sparse set of airfoil geometric information. As shown in Table \ref{tab:Points}, a comparison was carried out between the average performances of $\mathcal{D}^4_{05}$, $\mathcal{D}^4_{10}$ and $\mathcal{D}^4_{15}$ datasets, representing the upper and lower surface values at five, ten and fifteen fixed chordwise locations, respectively. The network architecture that was finalized from Table \ref{tab:Neurons} was chosen. It was observed that the neural network is powerful enough to capture the airfoil's aerodynamic characteristics even when minimum geometric information ($\mathcal{D}^4_{05}$) is provided, with the individual $R^2$ values coming out to be greater than 0.90. For the purpose of this research, $\mathcal{D}^4_{10}$ was selected since it gave slightly better average performance out of the three datasets.

\begin{figure}
  \centering
  \includegraphics[width=\linewidth]{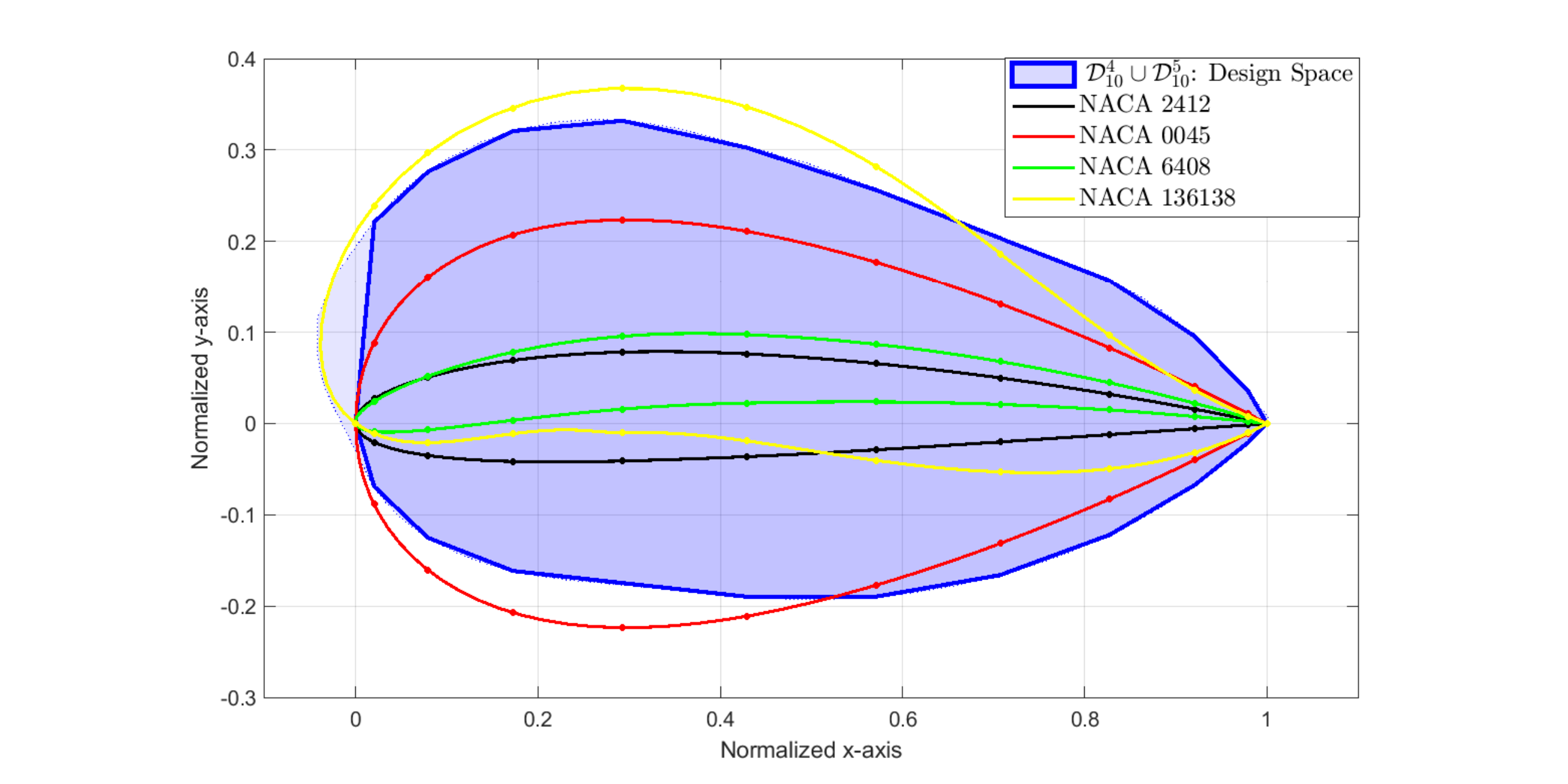}
  \caption{Design Space consisting of NACA 4- and 5-digit series}\label{fig:dspace}
\end{figure}

Furthermore, the authors deliberated on the neural network's capability of generalization. The idea: \emph{Will the network be able to predict the aerodynamic characteristics of 5-digit NACA series if it was trained on the 4-digit NACA series, and vice versa?}. The results outlined in Table \ref{tab:Dataset} seem to indicate that it is robust enough to do so. Upon analyzing, it was observed that the neural network learns the entire design space of geometric coordinates, shown in Figure \ref{fig:dspace}, as opposed to individual airfoil profiles. As long as the airfoil shape lies within this space, the network is likely to predict the characteristics with good levels of accuracy.

To further establish this hypothesis, a four-layered network was trained on combined dataset of NACA 4 and 5-digit series ($\mathcal{D}^4_{10} \cup \mathcal{D}^5_{10}$). Four airfoils (that were not present in the training dataset) were selected, namely NACA 0045, 2412, 6408 and 136138. These airfoils were passed through the trained network at Mach 0.25 and Reynolds Number 250000. It is important to highlight that the network was not explicitly trained on these flight conditions. However, these conditions were within the prescribed limits shown in Table \ref{tab:DataRanges}.

\begin{figure}
  \centering
  \includegraphics[width=\linewidth]{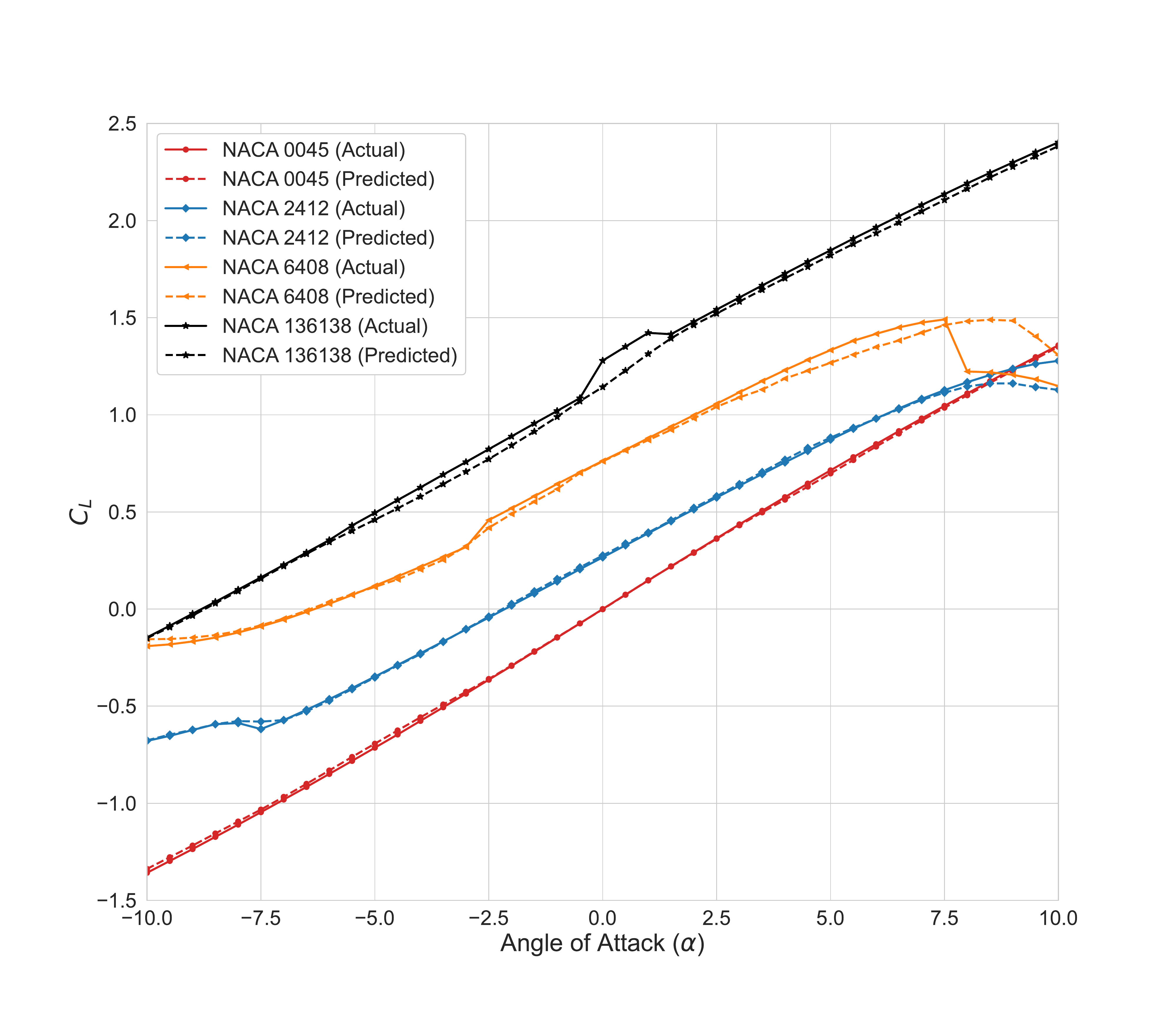}
  \caption{Lift Coefficient Prediction using ANN}\label{fig:CL}
\end{figure}

\begin{figure}
  \centering
  \includegraphics[width=\linewidth]{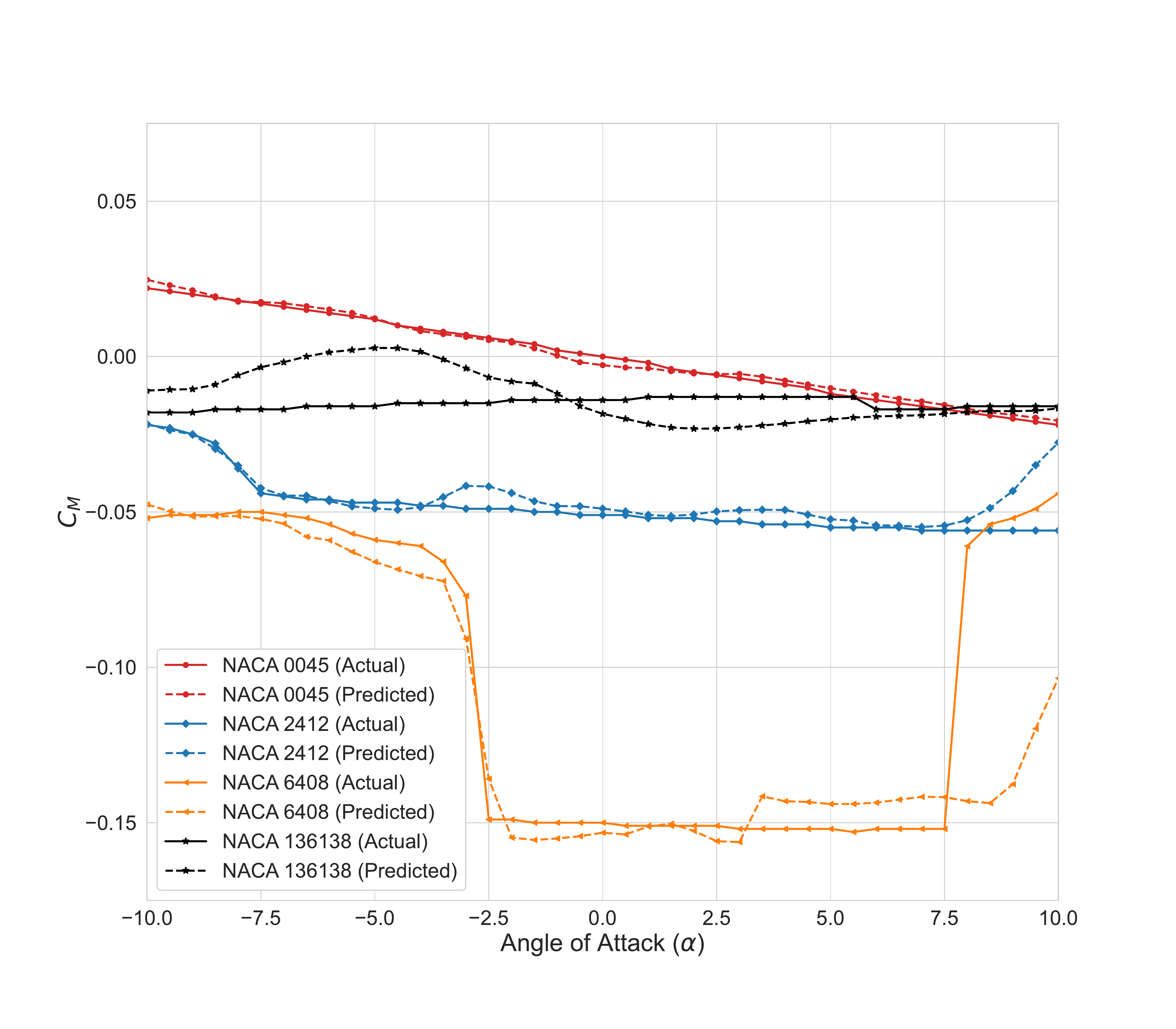}
  \caption{Pitching Coefficient Prediction using ANN}\label{fig:CM}
\end{figure}

\begin{figure}
  \centering
  \includegraphics[width=\linewidth]{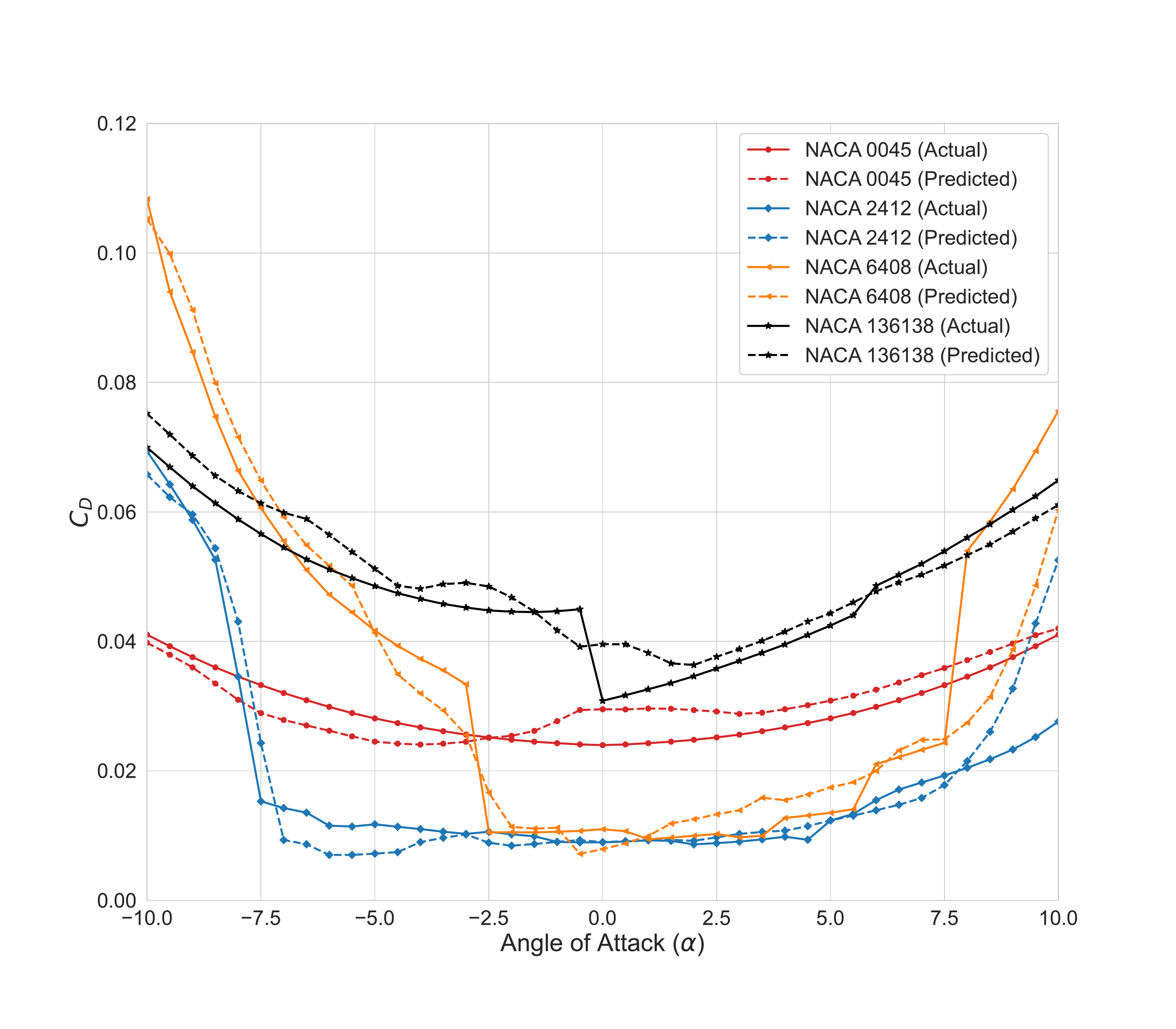}
  \caption{Drag Coefficient Prediction using ANN}\label{fig:CD}
\end{figure}

From Figure \ref{fig:dspace}, it can be seen that NACA 0045 and 136138 lie slightly out of the design space, while NACA 2412 and 6408 lie completely within it. Looking at $C_L$ vs. $\alpha$ graph (Figure \ref{fig:CL}), it is evident that the network adequately captures the lift distribution for each airfoil in the linear regime. However, the network fails to accurately capture stall regions of NACA 2412 and 6408, even when they both lie completely within the design space. In $C_m$ vs. $\alpha$ (Figure \ref{fig:CM}), it was observed that the network accurately predicted the $C_m$ vs. $\alpha$ trend, except for NACA 136138 which lies slightly outside the design space. An interesting result was observed for NACA 2412 and 6408. The network failed to accurately capture the original trend from 7.5 deg AoA onwards. This is the AoA from which both these airfoils start experiencing stall. Similar effect was noticed in the results of the drag polar. The $C_D$ vs $\alpha$ plot, or the drag polar (Figure \ref{fig:CD}) gave the lowest $R^2$ value amongst the three parameters of interest. For NACA 2412 and 6408, the network sufficiently predicts the outcome until the airfoils enter their stall regime, whereby the network diverges from the actual trend. For NACA 0045 and 136138, the network follows the trend adequately.

One possible reason that the network slightly diverges in the stall regime may be due to the limited range of the angle of attack provided in the dataset. Thus there are not many training examples that would help train the network in this regime. Furthermore, a good chunk of dataset consists of values of $C_L$ and $C_m$ that are linear with alpha, which may induce a form of bias that causes the entire network to become linear with AoA. The drag polar, however, is quadratic. In order to increase its $R^2$ value, one idea that the authors propose is to train the network to predict the parasitic drag ($C_{D_0}$) and lift induced drag coefficient ($K$), individually. The total $C_D$ can then be predicted using the drag polar equation ($C_D = C_{D_0} + K C_L^2 $).

\section{Conclusion}
In this research, an original idea of using sparse normalized 2D airfoil coordinates to predict lift, drag and moment coefficients using ANNs was presented. The training datasets consisting of NACA 4- and 5-digit airfoils, at different flight conditions, were generated using Javafoil. Multiple combinations of network architecture were trained, and their performance compared in order to identify the best architecture. Then, the network performance was investigated by varying the resolution of airfoil geometric coordinates. It was observed that the neural network is capable of capturing the aerodynamic characteristics even with sparse geometric information, unlike in CFD analysis. Another observation was that, the neural network learns the entire design space of geometric coordinates as opposed to individual airfoil profiles, due to which a network trained on NACA 4-digit series is robust enough to predict the results for NACA 5-digit series with reasonable levels of accuracy and vice versa. This study establishes that ANNs are capable of learning the aerodynamic characteristics easily and rapidly based on limited number of airfoil coordinates provided as input. This data driven method of predicting aerodynamic coefficients can be used in fully coupled aero-servo-elastic simulations at rapid pace with adequate accuracy; a task that would normally take massive computational resources if done through conventional numerical methods.

\bibliography{References}



\end{document}